\documentclass[pre,aps,twocolumn,showpacs,amsmath,amssymb]{revtex4}

\usepackage[dvips]{graphicx}
\usepackage{dcolumn}
\usepackage{bm}

\begin{document}

\preprint{}

\title{
Inhomogeneous substructures hidden in random networks
}

\author{Dong-Hee Kim and Hawoong Jeong}
\affiliation{Department of Physics, Korea Advanced Institute of Science and Technology, Daejeon 305-701, Korea}

\date{\today}

\begin{abstract}
We study the structure of the load-based spanning tree (LST) that 
carries the maximum weight of the Erd\"{o}s-R\'{e}nyi (ER) random network. 
The weight of an edge is given by the edge-betweenness centrality, 
the effective number of shortest paths through the edge. We find that 
the LSTs present very inhomogeneous structures in contrast to the 
homogeneous structures of the original networks. Moreover, it turns out that 
the structure of the LST changes dramatically as the edge density of
an ER network increases, from scale-free with a cutoff, scale-free, 
to a star-like topology. 
These would not be possible if the weights are randomly distributed,
which implies that topology of the shortest path 
is correlated in spite of the homogeneous topology of the random network.
\end{abstract}


\maketitle

Complex network theories have attracted much attention because of 
the usefulness to analyze diverse complex systems in the real world
\cite{Albert1,Dorogovtsev1}.
The most representative measure characterizing a network is 
the \textit{degree distribution}, $P(k)$, which indicates 
probability for a vertex to be directly connected to $k$ neighboring
vertices with edges. 
Specifically, while the power-law distribution, $P(k) \sim k^{-\gamma}$, 
is widely found in the most of real-world networks including technological, 
biological, and social networks, such as the Internet~\cite{Faloutsos1}, 
the World Wide Web~\cite{Albert2}, the metabolic networks~\cite{Jeong1}, 
the protein interaction networks~\cite{Jeong2}, and 
the coauthorship networks~\cite{Newman1},
there also exist homogeneous networks, such as the US highway network and 
the US power-grid network, which are explained by the bell-shaped 
and exponential degree distributions. Recently it has been claimed that 
an apparent scale-free network can be originated from a sampling result 
of a underlying homogeneous network~\cite{Clauset1}.  

While the degree distribution gives valuable knowledge of local 
structures of networks around us, it is also necessary to know 
global structures of networks to understand dynamics on the networks properly. 
The information transport between two vertices occurs 
along an optimal path connecting them, defined as the path minimizing the 
total cost~\cite{Braunstein1,Sreenivasan1,Buldyrev1}, 
which is usually determined by using the global knowledge of the network. 
For instance, full information of connections in the network 
is required to determine a shortest path defined as a path consisting of 
a minimum number of edges, which would be an optimal path 
if the costs of all edges are identical. 
The scale-free network has been revealed to have very 
inhomogeneous shortest path topology so that one can find extremely important 
vertices or edges that huge number of shortest paths are passing through, 
which has been supported by the power-law distribution of the betweenness 
centrality (BC) ~\cite{Goh1} and the existence of the transport skeleton 
structure~\cite{dhkim1} that makes it possible to understand the origin of 
the difference between the BC exponent classes of real-world 
networks~\cite{dhkim1,Goh1} and the universal properties of the fractal 
scaling~\cite{Goh2,Song1}. 
However, in the non-scale-free networks, even though it has been known that 
there are no such vertices or edges used heavily in the shortest paths,
the topology of the shortest paths has not been intensively studied so far. 

Our main interest is to find out how the topology of shortest paths 
is correlated with the network topology in the Erd\"{o}s-R\'{e}nyi (ER)
random network model~\cite{Erdos1}, where two arbitrary vertices 
in the network are randomly connected to each other by an edge 
with a given probability $p$, which gives the Poisson degree distribution. 
In order to systematically study the shortest path topology, 
it is necessary to treat the network as a weighted network 
in which the contribution of the shortest paths on each edge 
is assigned as the weight of the edge.
We use the \textit{edge-betweenness centrality} 
(edge-BC)~\cite{Freeman1,Girvan1,Newman2} to 
represent the contribution of the shortest paths on each edge of the network,
which is a convenient quantity counting the effective number of 
shortest paths through the edge and thereby gives the average 
traffic though the edges. 
The edge-BC of an edge $e_{ij}$ between vertices $i$ and $j$ is the total 
contribution of the edge on the shortest paths between all possible pairs 
of vertices, which is defined as follows:
\begin{equation}
b(e_{ij}) = \sum_{m \neq n} b(m,n;i,j)
= \sum_{m \neq n} \frac{c(m,n;i,j)}{c(m,n)},
\end{equation}
where $c(m,n;i,j)$ denotes the number of shortest paths 
from a vertex $m$ to $n$ through the edge $e_{ij}$, 
and $c(m,n)$ is the total number of shortest paths from $m$ to $n$.

In this weighted networks, one useful way to study the spatial 
correlation of the weight is to investigate the 
\textit{load-based spanning tree} (LST)~\cite{dhkim1}, 
which consists of a set of selected edges 
to maximize its total weights, which corresponds to the skeleton of the network
~\cite{dhkim1}. From the degree distribution of 
the LST, we can check whether the spatial distribution of the weights 
is correlated with the original network topology. If the weights are randomly 
distributed on the edges of the network, the degree distribution 
of the original network would be preserved in its LST~\cite{Szabo1}. 
On the other hand, if there exists significant 
topological correlation in the distribution of the weights on the edges,
the degree distribution is expected to show a large deviation from the 
Poisson distribution, the degree distribution of the ER network.  
In this paper, we investigate the structure properties of the LST of 
the ER model in this respect. We find that the LSTs show very inhomogeneous 
structures in contrast to the homogeneous structures of the original 
networks. It is found that the degree distribution of the LST shows 
rich characteristics depending on the connection probability 
and the size of the ER network, which turns out to be very different 
from the Poisson distribution.

\begin{figure}
\centering{\resizebox*{0.45\textwidth}{!}{\includegraphics{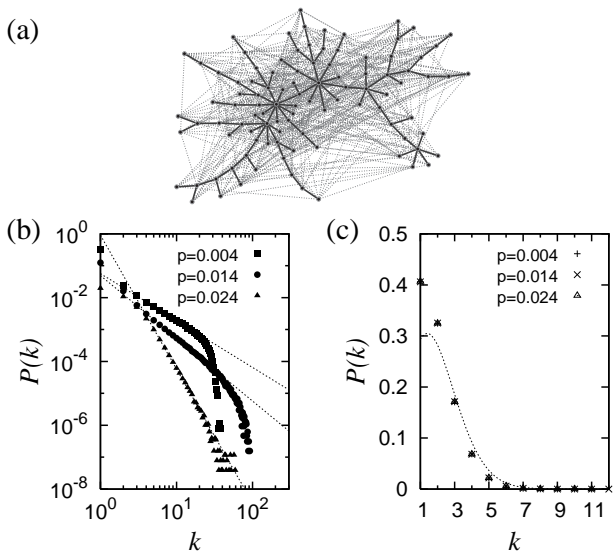}}}
\caption{
\label{fig:pk}
(a) An illustration of topologies of the LST (solid line) and the original 
ER network (dotted line) with $100$ vertices and $p=0.1$. 
(b) The degree distributions of the LSTs for various constructions of  
the ER networks with $N=4000$ vertices and connection probability 
of edges $p=0.004$,$0.014$,$0.024$. The degree distributions are guided by 
the curve-fits (dotted lines) to $P(k) \sim k^{-\gamma}$ with 
$\gamma=1.47$, $1.97$, $4.20$ for $p=0.004$, $0.014$, $0.024$, respectively.    
(c) The degree distribution of the LST when the weights are randomly 
redistributed on the edges. The Poisson degree distribution 
with average degree $2$ of trees (dotted line) is given 
for comparison. All data points are obtained from the averages 
over $256$ network ensembles.
}
\end{figure}

Figure~\ref{fig:pk}(a) shows an illustration of the inhomogeneous LST 
structure obtained from the ER network with $N=100$ vertices and $p=0.1$, 
in which the hub vertices having a significant number of degree are found.
The detailed characteristics of degrees can be found in 
Fig.~\ref{fig:pk}(b) which displays the degree distributions 
of the LSTs obtained from the ER networks generated with various 
connection probabilities. 
It is the most interesting feature that the right-skewed degree distributions 
are observed in the LSTs for the wide range of connection probability $p$ 
of the ER network whose degree distributions follow the narrow Poisson 
distribution.
For the examined LSTs, it is found that a power-law 
with a cutoff fits well to the degree distribution of the LST, where  
the exponent and the cutoff degree depend on $p$. 
The emergence of these inhomogeneous degree distributions in the LSTs 
indicates that there exist non-negligible correlations between 
the weights of neighboring edges sharing a common vertex at their ends, since 
a vertex shared by the edges having higher weights becomes a hub in the LST.
If we remove the correlation by the random redistribution~\cite{shuffle} 
of the weights on the edges, the degree distributions of the LSTs 
become the Poisson distribution as expected [see Fig.~\ref{fig:pk}(c)]. 
These imply that the shortest paths, 
which give the weights of its constituting edges, are not randomly 
distributed on the ER network but strongly correlated enough to 
generate an inhomogeneity in spite of the homogeneous 
topology of the ER network. 

The topological correlation of the shortest path that gives rise to 
the inhomogeneous LST structure can be specified by
the correlation between the degree of the vertex and the weights 
of the edges connected to the vertex. In order to find out more about 
the spatial correlation in the distribution of the weights, 
we measure the average weight rank $R_k$, an averaged value of weight-rank
$r$ over the edges attached to a vertex having degree $k$. 
The rank $r_{ij}$ of the edge between $i$ and $j$ is graded for 
its weight $w_{ij}$,
i.e., the largest weight gives $r=1$, the second largest weight gives $r=2$,
and so on. Mathematically $R_{k}$ is defined as follows:
\begin{equation}
R_k = \bigg\langle \frac{1}{|\mathbf{V}_k|}
\sum_{i \in \mathbf{V}_k} \frac{1}{k} \sum_{j}
r_{ij} a_{ij} \bigg\rangle ,
\end{equation}
where $\mathbf{V}_k$ and $|\mathbf{V}_k|$ are the set of vertices 
having degree $k$ and the number of those vertices, respectively, 
$a_{ij}$ is the adjacency matrix element; $a_{ij}=1$ 
if $i$ and $j$ is connected and $a_{ij}=0$ otherwise,
and $\langle \ldots \rangle$ denotes an average over network ensembles.  
Consequently, the small (large) value of $R$ indicates that a vertex
has the edge with high (low) ranks. The reason why we attach great 
importance to $R_k$ is that it gives an insight into how the 
structure of the network changes in the LST because the edges of the network 
are picked in the rank order for the LST. 
While the average value of the weights can give similar 
information, since the ER network has a rather homogeneous distribution of 
weight values, the average weight rank shows 
more clear changes depending on the spatial distribution of 
the weights as compared with the average value of the weights. 
In counting ranks, we do not admit a tie in ranks for
the edges with the same weights but assign ranks according to random
priority if the values of weights are the same. 
However, we note that averaging over the network ensembles 
removes the dependence on the counting method of the tie in ranks.

\begin{figure}
\centering{\resizebox*{0.45\textwidth}{!}{\includegraphics{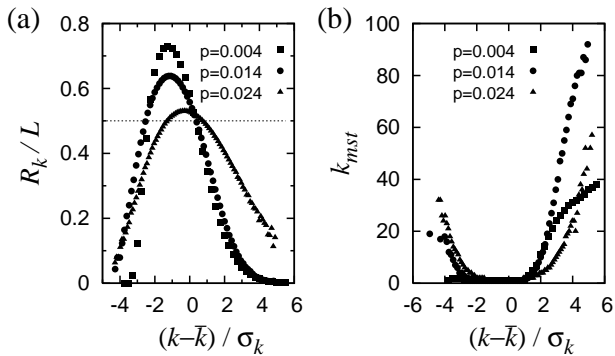}}}
\caption{
\label{fig:corr}
(a) The comparison of the average weight rank $R_k$ of the weighted ER 
network with that of the uncorrelated network (dotted line) generated by 
the random redistribution of weights on the edges.
(b) The degree correlation between the LST and the original ER network. 
ER networks with $N=4000$ vertices and connection probability 
of edges $p=0.004,0.014,0.024$ are examined. $\bar{k}$ and $\sigma_k^2$ 
denote the average and the variance of degrees in the original ER network, 
respectively. The data points are obtained from the averages 
over $256$ network ensembles. 
}
\end{figure}

In order to figure out how the weights are distributed on the network,  
we compare the average weight rank $R_{k}$ with that for the uncorrelated 
one in which the weights are randomly redistributed on the 
edges to remove the correlation between the edge and the weight on it. 
For instance, if $R_k$ is smaller than the value for uncorrelated weights,
the vertex with degree $k$ can be regarded to have an edge
with a higher weight as compared with the uncorrelated situation. 
This comparison gives an insight into the topological correlation 
of the weights on the network. 
From Fig.~\ref{fig:corr}(a), we find that the extreme vertices, which belong 
to the tail of either the smaller or larger degree part 
in the Poisson degree distribution, have significantly higher 
average weight ranks (small value of $R_k$)
than the average rank for the uncorrelated 
spatial distribution of the weights, given by $\sim L/2$, where 
$L$ is the total number of edges of the original network.
This indicates that these extreme vertices are more likely to 
preserve their degree in the LSTs. More directly, 
we also measure the degree correlation between the LST and its original 
network. In Fig.~\ref{fig:corr}(b), for the extreme vertices, 
it is confirmed that the degrees of the LST show strong correlations 
with the degrees of the original network. Consequently, 
the extreme vertices become hubs that constitute the heavy tail of 
the inhomogeneous degree distribution of the LST.

The structural homogeneity of the ER network varies with 
the edge density represented by using the connection 
probability $p$. As well as the structural properties,
the characteristics of the shortest path topology also depend 
on the edge density of the ER network.
Specifically, the change of the connection probability $p$ in
the ER network gives rise to the change of the degree 
distribution of its LST. At small values of $p$'s, as seen in 
Fig.~\ref{fig:pk}(b), the degree distribution 
of the LST can be characterized by the power-law exponent $\gamma$ 
and the average maximum degree $k_{max}$. 
Thus, we can monitor the structural change of the LST by looking at 
the dependence of $\gamma$ and $k_{max}$ on the connection probability $p$
and the network size $N$. 

\begin{figure}
\centering{\resizebox*{0.45\textwidth}{!}{\includegraphics{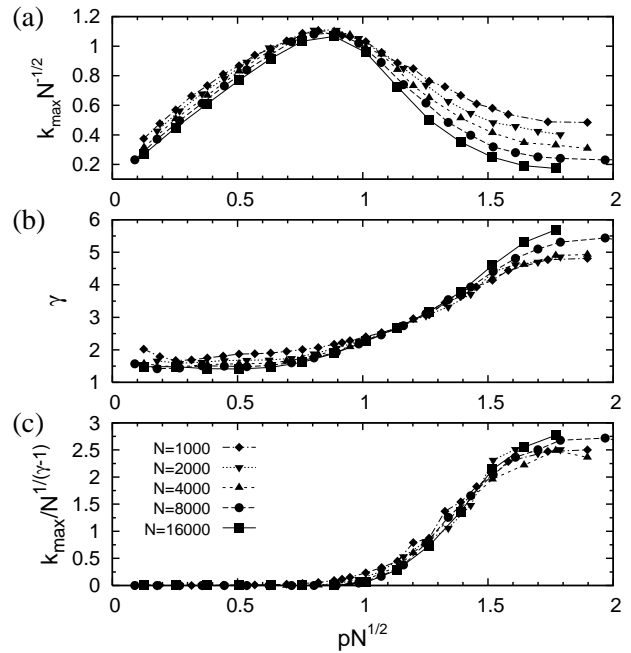}}}
\caption{
\label{fig:pdep}
The $p$ dependence of (a) the average maximum 
degree $k_{max}$ and (b) the power-law exponent $\gamma$ 
in the degree distribution of the LST. In (c), the comparison 
of $k_{max}$ with the natural cutoff $N^{1/(\gamma-1)}$ is given.
The data points are obtained from the averages over $256$ ensembles.
}
\end{figure}

As $p$ increases, $\gamma$ monotonically increases, but interestingly,
$k_{max}$ shows non-trivial behavior of decreasing even though 
$k_{max}$ has a chance to increase further as $p$ increases 
because a higher $p$ generates a larger 
average degree in the original ER network. To find a universal 
description of this non-trivial behavior of the degree distributions 
of the LSTs, we perform a size-scaling on data obtained from 
the ER networks of various sizes and connection probabilities 
and find out a common scaling feature;
$k_{max}$ monotonically increases to the maximum value of
$\sim N^{1/2}$ until $p$ reaches at $\sim N^{-1/2}$, then it 
decreases as $p$ increases further while $\gamma$ monotonically increases
[see Fig.~\ref{fig:pdep}].

This $p$ dependence of $\gamma$ and $k_{max}$ strongly 
implies that there exists a critical $p$ value at which a clear
scale-free degree distribution emerges by satisfying the balance between 
$\gamma$ and $k_{max}$. More precisely, because the natural cutoff 
$k_{cutoff}$ scales with $\sim N^{1/(\gamma -1)}$ in the finite sized 
network of $N$ vertices with power-law degree distribution 
$P(k)\sim k^{-\gamma}$~\cite{Cohen1,Dorogovtsev2}, 
we can test the \textit{scale-freeness}
of the degree distribution of the LST depending on $p$ by 
comparing $k_{max}$ with the natural cutoff. 
From the plot of $k_{max} / N^{1/(\gamma -1)}$ 
as a function of $pN^{1/2}$ shown in Fig.~\ref{fig:pdep}(c), 
we find that $k_{max}$ of the LST becomes comparable with the 
natural cutoff $N^{1/(\gamma-1)}$ when $p$ reaches at $N^{-1/2}$ 
so that the cutoff diminishes in the degree distribution, 
which leads to the emergence of the scale-free spanning tree.

\begin{figure}
\centering{\resizebox*{0.45\textwidth}{!}{\includegraphics{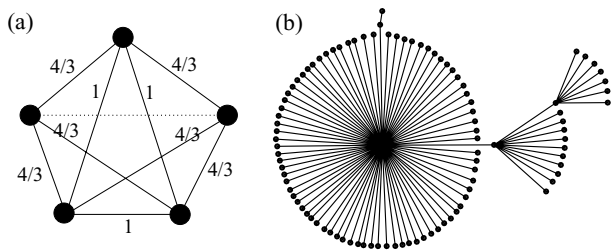}}}
\caption{
\label{fig:star}
(a) An example of the very dense network in which only one 
edges (dotted line) are eliminated from the fully connected network.     
The edge-BCs are given on the edges, which indicates that the edges of 
smaller degree vertices have more edge-BCs than the other edges.
(b) Typical topology of the LST for very dense ER network. The structure 
is obtained from the ER network with $N=100$ vertices and connection 
probability $p=0.95$. 
}
\end{figure}

The exponent $\gamma$ of the power-law degree distribution 
increases as $p$ increases in the small $p$ region as shown 
in Fig.~\ref{fig:pdep}, but the Poisson degree distribution 
of the original ER network is not recovered even if $p$ gets much larger. 
Interestingly, the average maximum degree $k_{max}$ increases 
when $p$ becomes far larger than $\sim N^{-1/2}$ \cite{comment1}. 
For intermediate values of $p$'s, the structures of LSTs show
very diverse degree distribution depending on the ensemble of the 
ER network generation since there can exist lots of edges 
that have same value of the edge-BC (weights), which gives 
large degeneracy in constructing LSTs. However,  
we find that a \textit{star-like} structure finally emerges 
in the LST [see Fig.~\ref{fig:star}] as the ER network gets very denser 
to be a nearly fully-connected network. 

In the limiting cases, this can be understood easily. Let us consider 
a nearly fully connected network in which only a single pair 
of vertices are not connected while all the other vertex pairs 
are directly connected by edges \cite{comment2}. 
In this case, the shortest path 
connecting these two vertices must pass through one of their common 
neighbor vertices, and consequently this shortest path gives 
additional contributions to the edge-BCs of the edges 
connected to these two vertices having smaller degrees 
[see Fig.~\ref{fig:star}(a)]. 
Therefore, in the edge selection for constructing LSTs,
the edges connected to the smaller degree vertices 
are chosen with higher priorities, which leads the emergence of 
the star-like LSTs.  

Finally, we note that the emergence of the scale-free spanning tree 
in the ER network also has been reported in the two recent works.
Clauset \textit{et al.} \cite{Clauset1} reported 
that the spanning tree constructed 
by using traceroutes from a single source has power-law degree 
distribution with cutoff. Kalisky \textit{et al.} \cite{Kalisky1} 
presented that the minimum spanning tree of the percolation cluster 
in the ER network shows the power-law degree distribution.
Together with these works, our findings indicate that the 
consideration of the transport pathways or the weights of the edges 
can drastically changes the original homogeneous topology of 
the ER network.

In conclusion, we have investigated the structural properties of 
the LST of the weighted ER network in which the weight of an edge 
is given by the edge-BC. We have found that the degree distribution of 
the LST shows very inhomogeneous distribution, which can be described
as power-law with a cutoff, scale-free, or star-like depending on 
the edge density of the ER network. The emergence of these inhomogeneous
degree distributions in the LST implies that the shortest paths 
are not homogeneously distributed on the ER network but spatially 
correlated in spite of the homogeneous topology of the ER network.  

\acknowledgments
This work was supported 
by Grant No. R14-2002-059-01002-0 from the KOSEF-ABRL Program (D.H.K.)
and by Grant No. R01-2005-000-1112-0 from the Basic Research Program
of the Korea Science \& Engineering Foundation (H.J.).


\begin{thebibliography}{10}
\bibitem{shuffle} We break up the pairs of the edges and their weights,
and randomly reassign the weights to the edges.
\bibitem{comment1} 
The value of $p$ at which $k_{max}$ starts to increase varies with 
the network size $N$, where we cannot find any size scaling relation or
universal feature.
\bibitem{comment2}
The extreme case of $p=1.0$ is out of our interest because every edge has
exactly same edge-BC and there is no inhomogeneity in the case 
of the perfectly fully connected network, in which any structure of 
an LST is possible, from star-like one to homogeneous one.
\end{thebibliography}
\end{document}